\documentstyle[preprint,aps,psfig]{revtex}
\tighten
\begin{document}
\preprint{Bicocca/FT  October 2000}

\title{A remark on the numerical validation of  triviality   for\\
 scalar field theories  using high-temperature expansions }

\author{P. Butera and M. Comi}

\address{Istituto Nazionale di Fisica Nucleare\\
Dipartimento di Fisica, Universit\`a di Milano-Bicocca\\
3 Piazza della Scienza,   20126 Milano, Italy \\E-mail:butera@mib.infn.it ; 
comi@mib.infn.it}

\maketitle

\narrowtext

\begin{abstract}
We suggest  a simple modification of the usual  procedures of analysis
 for the high-temperature (strong-coupling or hopping-parameter)
 expansions of the renormalized four-point coupling constant in  the 
$\vec \phi^4_4$ lattice scalar field theory. 
 As a result we can   more convincingly validate 
 numerically the triviality of the continuum limit taken 
 from the high temperature phase. 
\end{abstract}

  There has been a steady accumulation of suggestive numerical and analytical 
 evidence, but not yet 
 a complete rigorous proof that the continuum limit of the
 $N$-component $\vec \phi^4_4$ theory, (defined  on a lattice and with 
 ferromagnetic nearest-neighbor couplings), describes 
a  free\cite{sok,sokff,sym,calla} (or ``trivial") field theory.
 The basic clues of this paradoxical no-interaction property 
were clearly indicated almost half a century ago\cite{land},
 but  more stringent studies of this conjecture had to 
wait for the  developments
 of the Renormalization Group(RG) theory\cite{wk}.
 The modern rigorous analyses of Refs.\cite{sok,sokff,aize,fro,hara,shlo}
 have finally proved that the   $\vec \phi^4_d$ theory
 is non-trivial\cite{eck} in $d\leq 3 $  and trivial in $d\geq 5 $ dimensions, 
 at least for not too large values of $N$.
Triviality is expected to occur (and is supported by numerical 
 calculations) also in  $d= 4$ dimensions. 
 Since, however,   the 
 rigorous results in this direction are still partial,  
various routes\cite{galla,cons}
 to recover an interesting continuum theory have also been 
explored.

The  lattice Euclidean $\vec \phi^4_4$ theory with $O(N)$
 symmetry is defined by the action\cite{lw}  
\begin{equation}
S= \sum_{x}\Big\{-\beta \sum_{\mu}\vec\phi_x\cdot\vec\phi_{x+\mu} 
+ \vec\phi_x^2 + \lambda(\vec\phi_x^2 -1)^2\Big\}  
\label{action} \end{equation}

where $\vec\phi_x$ is a real $N$-component field at the  lattice site $x$,
  $\mu$ is the unit vector in the $\mu$  direction and $\lambda\geq 0 $.

For $\beta \uparrow \beta_c(N,\lambda)$, at fixed positive $\lambda$ 
 the model has a critical point 
where a second order transition occurs from a high-temperature(HT) 
paramagnetic phase  to a low-temperature ferromagnetic phase.
 In the $\lambda \rightarrow \infty$ limit, the model becomes to the lattice 
non-linear $O(N)$-symmetric 
$\sigma$-model or, equivalently,  the $N$-vector spin model.

 The construction of a continuum limit of the lattice theory 
is  reduced to the determination of its critical properties.
 Here we shall consider only the continuum limit taken from the HT phase. 
In the context of the RG theory a detailed description is obtained for  the 
asymptotic cutoff dependence of the correlation 
functions  in terms of the weak-coupling 
 expansion of  the theory's beta function.

If we set $\tau(N,\lambda) =1 - \beta/\beta_c(N,\lambda) $, 
 the perturbative RG
 theory  yields\cite{zinn}
 the following critical behavior   as $ \tau \downarrow 0$, 
 for the correlation length 
\begin{equation}
 \xi^2(\beta,N,\lambda) =
 A^2_{\xi}(N)\frac {\vert 
ln (\tau(N,\lambda))\vert^{G(N)}} {\tau(N,\lambda)}
\Big[1+ O\Big( ln(\vert ln(\tau)\vert)/ln(\tau)\Big)\Big]
\label{xi2}\end{equation}
 where $G(N)= \frac {N+2} {N+8}$. 

 The asymptotic  behavior of the susceptibility is completely similar 

\begin{equation}
 \chi(\beta,N,\lambda)= 
 A_{\chi}(N)\frac {\vert ln (\tau(N,\lambda))\vert^{G(N)}} {\tau(N,\lambda)}
\Big[1 + O\Big( ln(\vert ln(\tau)\vert)/ln(\tau)\Big) \Big] .
\label{chi}\end{equation}

The fourth
 derivative of the free energy at zero field $\chi_4(\beta,N,\lambda)$
 has the behavior 
\begin{equation}
 \chi_4(\beta,N,\lambda)= 
 A_{4}(N)\frac {\vert ln (\tau(N,\lambda))\vert^{4G(N)-1}} 
{\tau(N,\lambda)^4}
\Big[1 + O\Big( ln(\vert ln(\tau)\vert)/ln(\tau)\Big)\Big] .
\label{chi4}\end{equation}

In terms of $\chi$, $\xi^2$ and $\chi_4$,
the dimensionless renormalized  4-point coupling constant $g_r(N,\lambda)$
 is defined  by the critical value of the ratio 

\begin{equation}
 g_r(\beta,N,\lambda) =- \frac{ \chi_4(\beta,N,\lambda)} 
{ \xi^4(\beta,N,\lambda) \chi^2(\beta,N,\lambda)} 
\label{gierre}\end{equation}
 as $ \tau \downarrow 0$.
 
It can be shown that  $g_r(\beta,N,\lambda)$ is non negative\cite{lebo}
  for all $\beta$.
  If  $g_r(\beta,N,\lambda)$
vanishes 
as  $ \tau \downarrow 0$, the continuum  
limit of the lattice model  taken from the high-temperature phase describes 
  a (generalized)-free-field theory\cite{newman},
 namely a theory where the connected parts of the 
four-point and higher-point functions
 vanish. 

 As  $ \tau \downarrow 0$, the perturbative RG  yields the leading 
asymptotic behavior, 
 with a well specified universal amplitude\cite{lw} 
\begin{equation}
 g_r(\beta,N,\lambda) \approx \frac{c(N)} {\vert ln(\tau)\vert}\Big[ 1 +
 O\Big( ln(\vert ln(\tau)\vert)/ln(\tau)\Big)\Big] 
\label{glog}\end{equation}
 where $c(N)=2/b_1(N)$ and 
$b_1(N)=\frac {N+8} {48 \pi^2}$ is the first non-vanishing coefficient 
 of the beta-function.

 Therefore the perturbative RG theory  implies that 
$g_r(\beta,N,\lambda) \to 0$ as $\tau \downarrow 0$, namely that the continuum 
$\vec \phi^4_4$ model is  trivial.

Since the validity of the results in eqs.(\ref{xi2}), (\ref{chi}), 
(\ref{chi4}), (\ref{glog}) is based upon 
the (unwarranted) perturbative  determination of the beta function, 
it is   interesting, at least, to  try to 
confirm them within different approximation schemes.
 To this purpose, various HT or strong-coupling expansion 
 analyses\cite{wk,lw,gutt,gaunt,bakinc,vlad,bender} have been performed.
  Many extensive MonteCarlo (MC) lattice 
simulations\cite{car,wein,fox,dru,bern,frick,kim,grass,kenna}
have also been carried out  and progressively 
refined in the years, following 
the rapid evolution of computers and the improvement of algorithms and data
analysis. 
Until now, both the stochastic simulation 
 and the HT series studies
 have been generally carried out in a completely parallel  way. 
 For instance, in  the case of the 4d self-avoiding walk model 
(namely, the $\vec \phi^4_4$ model for $N=0$  and 
 $\lambda \rightarrow \infty$) a HT expansion  
of $\chi$ up to order $\beta^{13}$ 
 on the hypercubic lattice has been analyzed\cite{gutt}
  in order to detect the logarithmic factor predicted 
by the RG theory eq.(\ref{chi}) and  to estimate its
 exponent. 
For the same purpose, $\chi$ 
and $\xi^2$ have been measured\cite{car,grass} in 
 high precision MC simulations.
 Analogous studies\cite{gaunt,vlad} have been devoted to 
 the 4d Ising model (namely, 
the $N=1$ case for $\lambda \rightarrow \infty$)  using 
  series O($\beta^{17}$) for $\chi$ and $\chi_4$ on the hypercubic lattice.
 The MC simulations of Refs.
\cite{wein,fox,frick} have tried  to show directly the  
 consistency of the estimates of $ g_r(\beta,N,\lambda)$ with 
 the elusive asymptotic behavior eq.(\ref{glog}). 
 A somewhat different approach, based on the scaling properties of the 
 partition function zeroes in the complex $\beta$ plane, has been adopted 
 in the simulations of Ref.\cite{kenna}.
Moreover  various analytical or semianalytical approaches\cite{bard,grom}
 have also been pursued.  

 All of these non-perturbative calculations have given  results
 consistent, or at least not in contrast, 
 with the predicted critical behaviors of $\chi$, $\xi$, and $g_r$.
However, the cited computations are somewhat limited in their extent, 
since only the $N=0$ 
and $N=1$ cases for $ \lambda \rightarrow \infty$, and, the $N=1$ case for 
 finite $\lambda$ to order $\beta^{10}$   have
 been  considered\cite{bakinc} until now.
 Moreover,  it is a common experience  how  difficult it is  
 to uncover numerically a logarithmic behavior or, more generally, 
 a logarithmic correction to a power behavior. Indeed, 
as the computations proceed deeply 
into the asymptotic regime their reliability decreases and 
the uncertainties of their results often
reach almost the same order of magnitude as the effects 
that have to be revealed. 
In the case of the HT expansions,  we have also observed that, 
 the methods of Refs.\cite{gutt,vlad} which were very effective 
in the $N=0$ and $N=1$ cases, are not as  successful when $N>1$.

In this note, we do not present  new data, but reconsider
 the  HT expansions calculated through order $\beta^{14}$, 
more than a decade ago, by L\"uscher and Weisz\cite{lw} 
for $\chi$, $\xi^2$ and $\chi_4$ on the hypercubic 
 lattice.  They had produced and analysed these series to obtain a bound 
on the Higgs particle mass  as a consequence of the triviality 
of the scalar sector in the standard model. They used 
 from the outset also the assumption of validity of the perturbative RG 
and therefore avoided to place too much confidence in 
 the HT series within the critical regime.
 We make no such assumption, but rather we   suggest a slightly different 
 and  hopefully more  convincing way of analyzing the series, which takes
 advantage  of the specific smoothness features of the HT expansion approach
 and, in the end, also turns out to be completely consistent with 
the RG results. 
We study how accurately
 an obvious consequence of eq.(\ref{glog}), rather than the equation
 itself,  is confirmed by computations.
 
 We observe that eq.(\ref{glog}) implies that 

\begin{equation}
F(\beta,N,\lambda)= \tau(N,\lambda)\frac {d} {d\beta}
( \frac{1} { g_r(\beta,N,\lambda) }) = \frac{b_1(N)} {2} + 
O\Big( ln(\vert ln(\tau)\vert)/ln^3(\tau)\Big)
\label{effe} \end{equation}
 as $ \tau \downarrow 0$.

 In order to confirm triviality, one has then  to show at least, that 
  $F(\beta,N,\lambda)$  has a finite limit $\tilde F(N,\lambda)$, 
  as $ \tau \downarrow 0$. 
The analysis will be  even more compelling if  

 i)    $ \tilde F(N,\lambda)\approx \tilde F(N)$, 
namely if the quantity $\tilde F(N,\lambda)$   will appear  not to depend on 
$\lambda$, as required by universality,  

and if, moreover,  

ii) 
$\frac{2\tilde F(N)}{b_1(N)} \approx 1$, 
namely if, unlike in previous approaches, it is possible to show
 complete quantitative consistency between   the strong-coupling 
estimate of $g_r$, including  the universal amplitude $2/b_1(N)$, 
 and  the weak-coupling  RG prediction  eqs.(\ref{glog}),(\ref{effe}).

Since the HT series can be simply written\cite{bcv}
 as explicit functions of $N$,  we can easily repeat 
the analysis on a wide range of values of $N$ 
 and thus further corroborate this result. 

 Let us stress that the whole analysis cannot be easily 
 performed in the context of a MC 
simulation, whereas it is completely obvious in a HT series approach.

 The main results of our procedure 
 can  be summarized into a couple of figures.
 In Fig.1, for $N=4 $ , we have plotted the quantity 
$y=\frac{2\tilde F(\beta,N,\lambda)}{b_1(N)} $ 
versus the scaled variable $x=\beta/\beta_c(N,\lambda)$,  
in order to be able to compare the curves obtained 
for various fixed values of $\lambda$. 
 The values of $\beta_c(N,\lambda)$ used here have been estimated
 by an analysis of the 
susceptibility expansions.
 We have calculated $F(\beta,N,\lambda)$ 
by simply forming  Pad\'e Approximants (PA) of its HT expansion.
 For each value of $\lambda$, 
 we have plotted only the highest non-defective PA, namely 
the $[6/6]$ or the $[6/7]$ approximants, as appropriate.
 The other  approximants of sufficiently high order 
 have the same behavior and are not reported in the figure.
 As $x \rightarrow 1$, the various curves so obtained appear to tend to
 unity, independently of $\lambda$, within 
a good approximation, thus confirming i). 
 We expect that the residual small spread of the limiting values would be 
 significantly reduced  
 if we could further reduce the  uncertainties 
 in the determination of $\beta_c(N,\lambda)$ and 
 if we could devise approximants more accurately 
allowing for the  slowly vanishing 
 corrections to scaling indicated in eq.(\ref{effe}).
 Of course these improvements  are strictly related.

We have performed the above calculation for various values of $N$ and,  
 in Fig.2, we have plotted versus $N$ 
 the ratio $R(N)=\frac{2\tilde F(N)}{b_1(N)}$.
For each value of  $N$ the reported error reflects the spread 
of this quantity. 
 Within a   fair approximation,  $R(N)$ appears to be 
 unity over a wide range  of values of $N$, thus confirming ii). 

Therefore both graphs  indicate a good quantitative agreement with the 
asymptotic formula eq.(\ref{glog}), obtained within 
 the perturbative RG approach. These very general results are unlikely to be
 accidental and completely confirm the conventional expectations
 concerning triviality. 

The HT series we have used  in this first test are  definitely  too short to 
make a  more refined analysis possible. The favourable results, however, 
 suggest that this study should be resumed 
 as soon as our systematic work\cite{bctut} of  HT series extension by 
the linked-cluster method will make
 new  longer expansions available.  We do not expect  results
 qualitatively  different from this preliminary study, 
but  significant  quantitative improvements.

In conclusion, we have shown that a small modification 
of  current procedures of numerical analysis 
is sufficient to shift the emphasis from a difficult qualitative question,
 namely how accurately  an elusive logarithmic behavior is 
 reproduced by an approximation scheme of inevitably 
limited precision,
 to a more quantitative issue. Our  technique of 
analysis  for HT series is not more involved than the usual ones, while,
 for all values of $N$, it   seems to produce a more 
 convincing  numerical validation of the perturbative-RG triviality predictions
 within the  strong-coupling approach.

\begin{figure}[t]
\centerline{\hbox{
\psfig{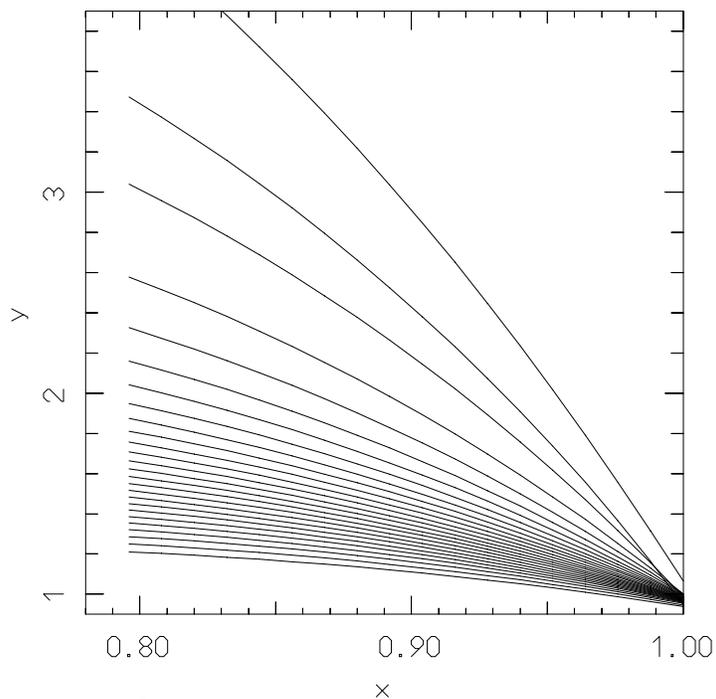}}}
\caption{The quantity 
$y=\frac{2\tilde F(\beta,N,\lambda)}{b_1(N)} $ 
versus the scaled variable $x=\beta/\beta_c(N,\lambda)$. We have taken
 $N=4$.
 Going from the top to the bottom of the figure, the various 
curves correspond to increasing values of $\lambda$ between 0 and $\infty$. }
\label{figura1klein} 
\end{figure}
\newpage
\begin{figure}[b]
\centerline{\hbox{
\psfig{figure=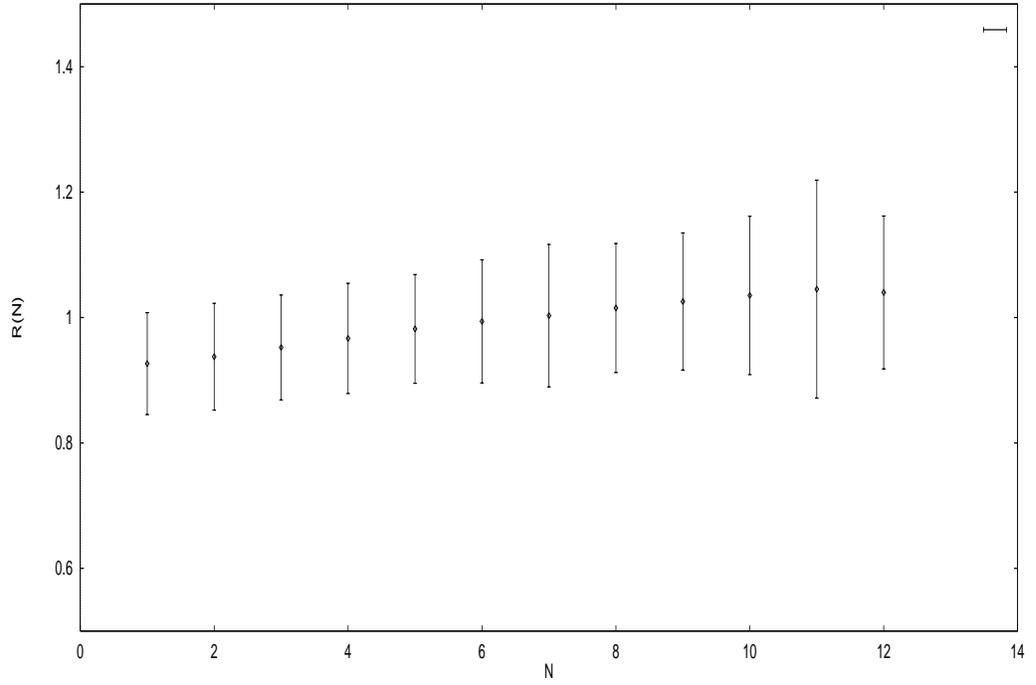,height=3.60in,width=4.2in}}}
\caption{The ratio $R(N)=\frac{2\tilde F(N)}{b_1(N)}$ versus $N$.}
\label{figura2klein} 
\end{figure}

\end{document}